\documentclass[11pt]{article}

\usepackage{amsmath,amsthm,verbatim,amssymb,amsfonts,amscd, graphicx}
\usepackage{graphics}
\usepackage{hyperref}
\usepackage{authblk}
\usepackage{lscape}
\usepackage{multirow}
\usepackage{float}

\theoremstyle{plain}
\numberwithin{equation}{section}

\date{}
\title{\textbf{Motion of Satellite under the Effect of Oblateness of Earth and Atmospheric Drag}}
\author[1]{Jaita Sharma\thanks{jaita.sharma@gmail.com}}
\author[1]{B. S. Ratanpal\thanks{bharatratanpal@gmail.com (Corresponding Author)}}
\author[2]{U. M. Pirzada\thanks{salmapirzada@yahoo.com}}
\author[1]{Vishant Shah\thanks{vishantmsu83@gmail.com}}
\affil[1]{\small{Department of Applied Mathematics, Faculty of Technology \& Engineering, The M. S. University of Baroda, Vadodara - 390 001, India}}
\affil[2]{\small{School of Science and Engineering, Navrachana University, Vadodara - 391 410, India}}

\begin{document}
\maketitle
\begin{abstract}
The equations governing motion of the satellite under the combined effects of oblate Earth and atmospheric drag have been studied, for a fixed initial position and three different initial velocities, till satellite collapses on Earth. In this study, we have considered exponential atmospheric density model and implemented R-K-Gill method. The minimum and maximum values of orbital elements and their variation over a time for different initial velocities have been reported.
\end{abstract}


\section{Introduction}
The study of motion of the satellite and its life span is the topic of interest of many researchers over the past few decades. When the orbit of the satellite is low Earth orbit (LEO), the perturbation due to oblateness of Earth and atmospheric drag plays very important role. Various analytic, semi-analytic and numerical techniques are adopted for solving perturbed equations of
motion. Raj\cite{Raj} extensively studied the motion of satellite under the oblateness of Earth and also by 
considering atmospheric drag. They solved the equations of motion by applying KS transformations \cite{SS}. King-Hele\cite{King}
solved the equations of the motion of a satellite analytically by considering oblateness of Earth. The motion of satellite 
in the terrestrial upper atmosphere was studied by Sehnal\cite{Sehnal}. Knowles \textit{et. al.}\cite{KPTN} analyzed the effect of geomagnetic strom's driven
by solar eruption on upper atmosphere of Earth and its effect on motion of satellite. The dynamics of satellite motion 
around the oblate Earth using rotating frame were developed by Yan and Kapila\cite{YK}. The Hamilton equations for the motion of satellite 
under the Earth's oblateness and atmospheric drag were derived and solved using canonical transformation by Khalil\cite{Khalil}.
Bezdv\v{e}k and Vokrouhlick\'{y}\cite{BV} presented a semi-analytic theory for long-term dynamics of a low Earth orbit of artificial satellites, they
considered both oblateness of Earth and atmospheric drag. Some statistical measures were used by them to compare the
observations over the computer efficiency. The resonance in satellite motion under air drag was studied by Bhardwaj and Sethi\cite{BS}. Hassan \textit{et. al.}\cite{HHB} tried to find a solution of equations governing motion of artificial satellites under the effect of an oblateness of
Earth by using KS variables. The authors then applied Picard's iterative method to find the solution. The algorithm is 
prescribed by the authors depends on initial guess solution. The differential equations governing relative motion of the
satellite under the oblateness of Earth and atmospheric drag were derived and solved by Chen and Jing\cite{CJ}, the wide application of
their work is in satellite attitude control and orbital maneuver for inter-planetary missions. The satellite rotational
dynamics was studied and simulated by Lee \textit{et. al.}\cite{LSSC}, they used Lie group variational integrator approach. Reid and Misra\cite{RM} studied the effect of
aerodynamic forces on the formation flight of satellite. The analytic solution in terms of Keplerian angular elements of
satellite orbit under atmospheric drag was studied by Xu and Chen\cite{XTCY}. Al-Bermani \textit{et. al.}\cite{AAAB} investigated the effect of atmospheric drag and
zonal harmonic $J_2$ for the near Earth orbit satellite namely $ Cosmos1484$. The analytic solution of motion of satellite
by considering combined effect of Earth's gravity and air drag was found by Delhaise\cite{Delhaise} using Lie transformations. 
Aghav and Gangal\cite{AG} designed and simplified the orbit determination algorithm for low Earth orbit navigation.

In this paper we have used R-K Gill method to solve the equations of motion of satellite under the influence of oblateness of Earth and atmospheric
drag in the low Earth orbit. We have analyzed results for $1-$day, $1-$month, $6-$months and till satellite collapses on Earth,
by considering fixed initial position and three different initial velocities. The orbital elements have been computed for the above mentioned period. We have considered the initial velocities in such a way that satellite will remain in low Earth orbit.

The paper is organized as follows: section 2 describes the model. Solution of equations governing motion of the satellite under
oblateness of Earth and atmospheric drag and calculation of orbital elements are reported in section 3. Section 4 contains 
discussion and concluding remarks.

\section{The Model}
\noindent The equation of motion of satellite without any additional perturbing force other than gravitational force between 
Earth and satellite is given by
\begin{equation}
 \ddot{\vec{r}}= - \frac{\mu}{r^3}\vec{r} \label{EMW},
\end{equation}
where $\mu= GM$, $G$ is gravitational constant and $M$ is mass of Earth. In the presence of perturbation, additional perturbing acceleration must be added on the right side of equation (\ref{EMW}). Since we are considering perturbation due to
oblateness of Earth and perturbation due to atmospheric drag, the equation of motion can be written as
\begin{equation}
 \ddot{\vec{r}}= - \frac{\mu}{r^3}\vec{r}+\vec{a}_O+\vec{a}_A \label{EMP},
\end{equation}
where $\vec{a}_O$ is acceleration due to oblateness of Earth and $\vec{a}_A$ is acceleration due to atmospheric drag.
The second order equation (\ref{EMP}) can be written as following set of two first order differential equations
\begin{align}
 \begin{split}
\dot{\vec{r}}&=\vec{v},\\
\dot{\vec{v}}&= - \frac{\mu}{r^3}\vec{r}+\vec{a}_O+\vec{a}_A.
 \end{split}\label{EMPFO}
\end{align}

In the Cartesian co-ordinate system the system of equations (\ref{EMPFO}) takes the form,
\begin{align}
 \begin{split}
  \dot{x}&=v_{x},\\
  \dot{y}&=v_{y},\\
  \dot{z}&=v_{z},\\
  \dot{v_x}&=-\frac{\mu x}{r^3}+\vec{a}_{O_x}+\vec{a}_{A_x},\\
  \dot{v_y}&=-\frac{\mu y}{r^3}+\vec{a}_{O_y}+\vec{a}_{A_y},\\
  \dot{v_z}&=-\frac{\mu z}{r^3}+\vec{a}_{O_z}+\vec{a}_{A_z},
 \end{split}\label{EMPSix1}
\end{align}
where $\vec{a}_{O_x},\vec{a}_{O_y}$ and $\vec{a}_{O_z}$ are components of acceleration due to oblateness of Earth in the 
direction $x,y$ and $z$ axis respectively and $\vec{a}_{A_x},\vec{a}_{A_y}$ and $\vec{a}_{A_z}$ are components of acceleration
due to atmospheric drag in $x,y$ and $z$ axis respectively.\\
The Earth's gravitational potential can be modeled in terms of zonal harmonics Battin(\cite{Battin}).  In the expression the value of
$J_2$ zonal coefficient is $400$ times higher than other $J_n$ zonal coefficient, $n\geq 3$. Hence we consider only
$J_2$ into account. If these higher order zonal coefficients are neglected and taking the gradient of scalar potential
function then the components of acceleration due to oblateness of Earth in the direction of $x, y$ and $z$ direction
respectively are,
\begin{align}
\begin{split}
 \vec{a}_{O_x}&=-\frac{3\mu R^2 J_2 x (x^2+y^2-4z^2)}{2r^7},\\
 \vec{a}_{O_y}&=-\frac{3\mu R^2 J_2 y (x^2+y^2-4z^2)}{2r^7},\\
 \vec{a}_{O_z}&=-\frac{3\mu R^2 J_2 z (3x^2+3y^2-2z^2)}{2r^7},
\end{split}\label{aO}
 \end{align}
where $R= 6378.1363\ km$ is radius of Earth, $\mu =GM= 398600.436233\ km^3/sec^2$ and $J_2= 1082.63 \times 10^{-6}$.\\
The acceleration due to atmospheric density is given by 
\begin{equation}
 \vec{a}_A=-\frac{1}{2} \rho\frac{ C_D A}{m} |\vec v_r|\vec{v_r}, \label{aAv}
\end{equation}
where $\rho$ is atmospheric density, $C_D$ is drag coefficient, $A$ is cross sectional area of the satellite perpendicular 
to velocity vector, $m$ is mass of satellite and $\vec{v_r}$ is satellite velocity vector relative to an atmosphere.\\
We take the simple exponential atmospheric model for which atmospheric density given by,
\begin{equation}
 \rho =\rho_{pa} e^{\left[\frac{(r_{pa} -r)}{H}\right]}, \label{rho} 
\end{equation}
where $\rho_{pa}$ is the density at initial perigee point, $r_{pa}$ is the initial distance of satellite from Earth's
surface, $r=|\vec{r}|$ and $H$ is scale height. The ratio $B^*= \frac{C_D A}{m}$ is called the Ballistic coefficient.

We assume that the atmosphere rotates at the same angular speed as Earth. With this assumption the relative velocity vector
is given by Wiesel\cite{Wiesel}
\begin{equation}
 \vec{v_r}=\vec{v}-\vec{\omega}\times \vec{r}, \label{vrvec}
\end{equation}
where, $\vec{\omega}$ is the inertial rotation vector of the Earth given by 
\begin{equation}
 \vec{\omega} = \omega_{e} \begin{bmatrix} 0\\0\\1 \end{bmatrix}, \label{omega}
\end{equation}
where, $\omega_e=7.292115486 \times 10^{-5} \ rad/sec$. The cross product of the (\ref{vrvec}) and (\ref{omega}) gives three
components of the relative velocity vector as
\begin{equation}
 \vec{v_r}=\begin{bmatrix} v_x+\omega_e r_y\\v_y-\omega_e r_x\\v_z\end{bmatrix}. \label{vrvec1}
\end{equation}
Substituting (\ref{rho}), (\ref{vrvec1}) and $B^*$ in (\ref{aAv}), we get the components of acceleration due to atmospheric drag
in the direction of $x,y$ and $z$ axis respectively as
\begin{align}
 \begin{split}
  a_{A_x}&= - \frac{\rho_{pa} e^{[\frac{r_{pa}-r}{H}]} \sqrt{(v_x+\omega_e r_y)^2+(vy-\omega_e r_x)^2+v^{2}_z}\; (v_x+\omega_e r_y)B^*}{2},\\
  a_{A_y}&= - \frac{\rho_{pa} e^{[\frac{r_{pa}-r}{H}]} \sqrt{(v_x+\omega_e r_y)^2+(vy-\omega_e r_x)^2+v^{2}_z}\; (vy-\omega_e r_x)B^*}{2},\\
  a_{A_z}&= - \frac{\rho_{pa} e^{[\frac{r_{pa}-r}{H}]} \sqrt{(v_x+\omega_e r_y)^2+(vy-\omega_e r_x)^2+v^{2}_z}\; v_z B^*}{2}.
 \end{split} \label{aA}
\end{align}
Substituting (\ref{aO}) and (\ref{aA}) in (\ref{EMPSix1}), we get equations of motion of satellite under oblateness of Earth
and atmospheric drag as
\begin{footnotesize}
\begin{align}
 \begin{split}
  \dot{x}&=v_{x},\\
  \dot{y}&=v_{y},\\
  \dot{z}&=v_{z},\\
  \dot{v_x}&=-\frac{\mu x}{r^3}-\frac{3\mu R^2 J_2 x (x^2+y^2-4z^2)}{2r^7}-\frac{\rho_{pa} e^{[\frac{r_{pa}-r}{H}]} \sqrt{(v_x+\omega_e r_y)^2+(vy-\omega_e r_x)^2+v^{2}_z}\; (v_x+\omega_e r_y)B^*}{2},\\
  \dot{v_y}&=-\frac{\mu y}{r^3}-\frac{3\mu R^2 J_2 y (x^2+y^2-4z^2)}{2r^7}-\frac{\rho_{pa} e^{[\frac{r_{pa}-r}{H}]} \sqrt{(v_x+\omega_e r_y)^2+(vy-\omega_e r_x)^2+v^{2}_z}\; (vy-\omega_e r_x)B^*}{2},\\
  \dot{v_z}&=-\frac{\mu z}{r^3}-\frac{3\mu R^2 J_2 z (3x^2+3y^2-2z^2)}{2r^7}-\frac{\rho_{pa} e^{[\frac{r_{pa}-r}{H}]} \sqrt{(v_x+\omega_e r_y)^2+(vy-\omega_e r_x)^2+v^{2}_z}\; v_z B^*}{2}.
   \end{split}\label{EMPSix2}
\end{align}
\end{footnotesize}
\pagebreak
\\
For the exponential atmospheric model the scale height ($H$) and $\rho_{pa}$ can be computed from the table-1,  Vallado(\cite{Vallado}).
\begin{table}{H}
\centering
\caption{Density at Initial Perigee Point and Scale Height}
\label{tab:1}
\centering\small
\begin{tabular}{|c|c|c|}
\hline
 \textbf{$ r_{pa}\;(km) $} & \textbf{$ \rho_{pa}\;(kg/m^3) $} & \textbf{$ H\;(km) $} \\

\hline
	$0$	&	$1.225$			&	$7.249$\\\hline
	$25$	&	$3.899\times10^{-2}$	&	$6.349$\\\hline
	$30$	&	$1.774\times10^{-2}$	&	$6.682$\\\hline
	$40$	&	$3.972\times10^{-3}$	&	$7.554$\\\hline
	$50$	&	$1.057\times10^{-3}$	&	$8.382$\\\hline
	$60$	&	$3.206\times10^{-4}$	&	$7.714$\\\hline
	$70$	&	$8.770\times10^{-5}$	&	$6.549$\\\hline
	$80$	&	$1.905\times10^{-5}$	&	$5.799$\\\hline
	$90$	&	$3.396\times10^{-6}$	&	$5.382$\\\hline
	$100$	&	$5.297\times10^{-7}$	&	$5.877$\\\hline
	$110$	&	$9.661\times10^{-8}$	&	$7.263$\\\hline
	$120$	&	$2.438\times10^{-8}$	&	$9.473$\\\hline
	$130$	&	$8.484\times10^{-9}$	&	$12.636$\\\hline
	$140$	&	$3.845\times10^{-9}$	&	$16.149$\\\hline
	$150$	&	$2.070\times10^{-9}$	&	$22.523$\\\hline
	$180$	&	$5.464\times10^{-10}$	&	$29.740$\\\hline
	$200$	&	$2.784\times10^{-10}$	&	$37.105$\\\hline
	$250$	&	$7.248\times10^{-11}$	&	$45.546$\\\hline
	$300$	&	$2.418\times10^{-11}$	&	$53.628$\\\hline
	$350$	&	$9.518\times10^{-12}$	&	$53.298$\\\hline
	$400$	&	$3.725\times10^{-12}$	&	$58.515$\\\hline
	$450$	&	$1.585\times10^{-12}$	&	$60.828$\\\hline
	$500$	&	$6.967\times10^{-13}$	&	$63.822$\\\hline
	$600$	&	$1.454\times10^{-13}$	&	$71.835$\\\hline
	$700$	&	$3.614\times10^{-14}$	&	$88.667$\\\hline
	$800$	&	$1.170\times10^{-14}$	&	$124.64$\\\hline
	$900$	&	$5.245\times10^{-15}$	&	$181.05$\\\hline
	$1000$	&	$3.019\times10^{-15}$	&	$268.00$\\\hline
\end{tabular}\normalsize
\end{table}
\section{Solution and Calculation of Orbital Elements}
\noindent Using R-K Gill method, we have solved the differential equations (\ref{EMPSix2}) by fixing the initial position and varying the initial velocity.

\noindent We have fixed the initial position $\vec{r}_{0}=[0, -5888.9727, -3400]$ in kilometers, ballistic coefficient $B^{*}=0.095\;m^{2}/kg$ and three different initial
velocities are considered as (i) $\vec{v}_{0}=[7.6, 0, 0]$, (ii) $\vec{v}_{0}=[7.7, 0, 0]$ and (iii) $\vec{v}_{0}=[7.8, 0, 0]$ in $km/sec^{2}$. For each of these three initial velocities which lead to low Earth orbit we have solved the system of differential equations (\ref{EMPSix2}) using R-K Gill method.
We have analyzed each of three cases till satellite hits on the Earth. We have obtained the intermidiate values of scale height $(H)$ and $\rho_{pa}$ by applying interpolation on values described in Table-1. The minimum and maximum values of orbital elements for each of these three initial velocities over different time periods are shown in table-2, table-3 and table-4 respectively.
\pagebreak

\begin{table}[h]
\centering\tiny
\caption{$\vec{r}_{0}=[0, -5888.9727, -3400]\;km;\;\vec{v}_{0}=[7.6, 0, 0]\;km/sec;B^{*}=0.096\;m^{2}/kg;$ Satellite collapses after 3222.645833333333 days.}
\label{tab:2}
\begin{tabular}{|c|c|c|c|c|c|c|c|c|}
\hline
\multirow{2}{*}{\textbf{Orbital Elements}} & \multicolumn{2}{c|}{\textbf{1 Day}}            & \multicolumn{2}{c|}{\textbf{30 Days}}          & \multicolumn{2}{c|}{\textbf{180 Days}}         & \multicolumn{2}{c|}{\textbf{3222 Days}}         \\ \cline{2-9} 
                                           & \textbf{Min}           & \textbf{Max}          & \textbf{Min}           & \textbf{Max}          & \textbf{Min}           & \textbf{Max}          & \textbf{Min}           & \textbf{Max}          \\ \hline
\textbf{$a$}                               & $6.7019\times 10^{3}$  & $6.7070\times 10^{3}$ & $6.6999\times 10^{3}$  & $6.7070\times 10^{3}$ & $6.6896\times 10^{3}$  & $6.7070\times 10^{3}$ & $6.4476\times 10^{3}$  & $6.7070\times 10^{3}$ \\ \hline
\textbf{$e$}                               & $0.0144$               & $0.0161$              & $0.0140$               & $0.0165$              & $0.0137$               & $0.0165$              & $0.0089$               & $0.0165$              \\ \hline
\textbf{$i$}                               & $0.5236$               & $0.5242$              & $0.5236$               & $0.5242$              & $0.5236$               & $0.5242$              & $0.5235$               & $0.5242$              \\ \hline
\textbf{$\Omega$}                          & $0$                    & $6.2829$              & $0$                    & $6.2829$              & $0$                    & $6.2830$              & $0$                    & $6.2832$              \\ \hline
\textbf{$\omega$}                          & $1.4938$               & $1.8496$              & $7.1754\times 10^{-4}$ & $6.2829$              & $8.4416\times 10^{-5}$ & $6.2831$              & $1.7420\times 10^{-6}$ & $6.2832$              \\ \hline
\textbf{f}                                 & $0.0359$ & $6.2820$              & $5.7651\times 10^{-4}$ & $6.2828$              & $5.3125\times 10^{-6}$ & $6.2831$              & $5.3125\times 10^{-6}$ & $6.2832$              \\ \hline
\end{tabular}\normalsize
\end{table}

\begin{table}[h]
\centering\tiny
\caption{$\vec{r}_{0}=[0, -5888.9727, -3400]\;km;\;\vec{v}_{0}=[7.7, 0, 0]\;km/sec;B^{*}=0.096\;m^{2}/kg;$ Satellite collapses after 6523.695833333334 days.}
\label{tab:3}
\begin{tabular}{|c|c|c|c|c|c|c|c|c|}
\hline
\multirow{2}{*}{\textbf{Orbital Elements}} & \multicolumn{2}{c|}{\textbf{1 Day}}            & \multicolumn{2}{c|}{\textbf{30 Days}}          & \multicolumn{2}{c|}{\textbf{180 Days}}         & \multicolumn{2}{c|}{\textbf{6523 Days}}         \\ \cline{2-9} 
                                           & \textbf{Min}           & \textbf{Max}          & \textbf{Min}           & \textbf{Max}          & \textbf{Min}           & \textbf{Max}          & \textbf{Min}           & \textbf{Max}          \\ \hline
\textbf{$a$}                               & $6.8787\times 10^{3}$  & $6.8837\times 10^{3}$ & $6.8772\times 10^{3}$  & $6.8837\times 10^{3}$ & $6.8690\times 10^{3}$  & $6.8837\times 10^{3}$ & $6.4156\times 10^{3}$  & $6.8837\times 10^{3}$ \\ \hline
\textbf{$e$}                               & $0.0101$               & $0.0117$              & $0.0097$               & $0.0121$              & $0.0096$               & $0.0121$              & $0.0039$               & $0.0121$              \\ \hline
\textbf{$i$}                               & $0.5236$               & $0.5242$              & $0.5236$               & $0.5242$              & $0.5236$               & $0.5242$              & $0.5234$               & $0.5242$              \\ \hline
\textbf{$\Omega$}                          & $0$                    & $6.2829$              & $0$                    & $6.2829$              & $0$                    & $6.2831$              & $0$                    & $6.2832$              \\ \hline
\textbf{$\omega$}                          & $4.6148$               & $4.9996$              & $1.4771\times 10^{-4}$ & $6.2825$              & $8.3173\times 10^{-5}$ & $6.2832$              & $4.1989\times 10^{-7}$ & $6.2832$              \\ \hline
\textbf{f}                                 & $0$ 		    & $6.2727$              & $0$ 		     & $6.2827$              & $0$ 		      & $6.2831$              & $0$ 		       & $6.2832$              \\ \hline
\end{tabular}\normalsize
\end{table}

\begin{table}[h]
\centering\tiny
\caption{$\vec{r}_{0}=[0, -5888.9727, -3400]\;km;\;\vec{v}_{0}=[7.8, 0, 0]\;km/sec;B^{*}=0.096\;m^{2}/kg;$ Satellite collapses after 9150.111111111111 days.}
\label{tab:4}
\begin{tabular}{|c|c|c|c|c|c|c|c|c|}
\hline
\multirow{2}{*}{\textbf{Orbital Elements}} & \multicolumn{2}{c|}{\textbf{1 Day}}            & \multicolumn{2}{c|}{\textbf{30 Days}}          & \multicolumn{2}{c|}{\textbf{180 Days}}         & \multicolumn{2}{c|}{\textbf{9150 Days}}         \\ \cline{2-9} 
                                           & \textbf{Min}           & \textbf{Max}          & \textbf{Min}           & \textbf{Max}          & \textbf{Min}           & \textbf{Max}          & \textbf{Min}           & \textbf{Max}          \\ \hline
\textbf{$a$}                               & $7.0676\times 10^{3}$  & $7.0726\times 10^{3}$ & $7.0663\times 10^{3}$  & $7.0728\times 10^{3}$ & $7.0589\times 10^{3}$  & $7.0728\times 10^{3}$ & $6.4832\times 10^{3}$  & $7.0728\times 10^{3}$ \\ \hline
\textbf{$e$}                               & $0.0366$               & $0.0381$              & $0.0362$               & $0.0385$              & $0.0359$               & $0.0385$              & $0.0143$               & $0.0385$              \\ \hline
\textbf{$i$}                               & $0.5236$               & $0.5242$              & $0.5236$               & $0.5242$              & $0.5236$               & $0.5242$              & $0.5234$               & $0.5242$              \\ \hline
\textbf{$\Omega$}                          & $0$                    & $6.2829$              & $0$                    & $6.2829$              & $0$                    & $6.2831$              & $0$                    & $6.2832$              \\ \hline
\textbf{$\omega$}                          & $4.6914$               & $4.9048$              & $1.7723\times 10^{-4}$ & $6.2828$              & $7.4474\times 10^{-5}$ & $6.2832$              & $3.6317\times 10^{-7}$ & $6.2832$              \\ \hline
\textbf{f}                                 & $0$ 		    & $6.2770$              & $0$ 		     & $6.2831$              & $0$ 		      & $6.2832$              & $0$ 		       & $6.2832$              \\ \hline
\end{tabular}\normalsize
\end{table}
\noindent The orbit of the satellite for satellite with initial position $\vec{r}_{0}=[0, -5888.9727, -3400]\;km$, 
initial velocity $\vec{v}_{0}=[7.8, 0, 0]\;km/sec$ and ballistic coefficient $B^{*}=0.096\;m^{2}/kg;$ for 1 day, 3 days and 7 days are shown in figure 1 respectively
from left to right.\\
\pagebreak

\begin{figure}[ht]
\begin{center}
\begin{tabular}{ccc}
\includegraphics[width=6cm,height=6cm]{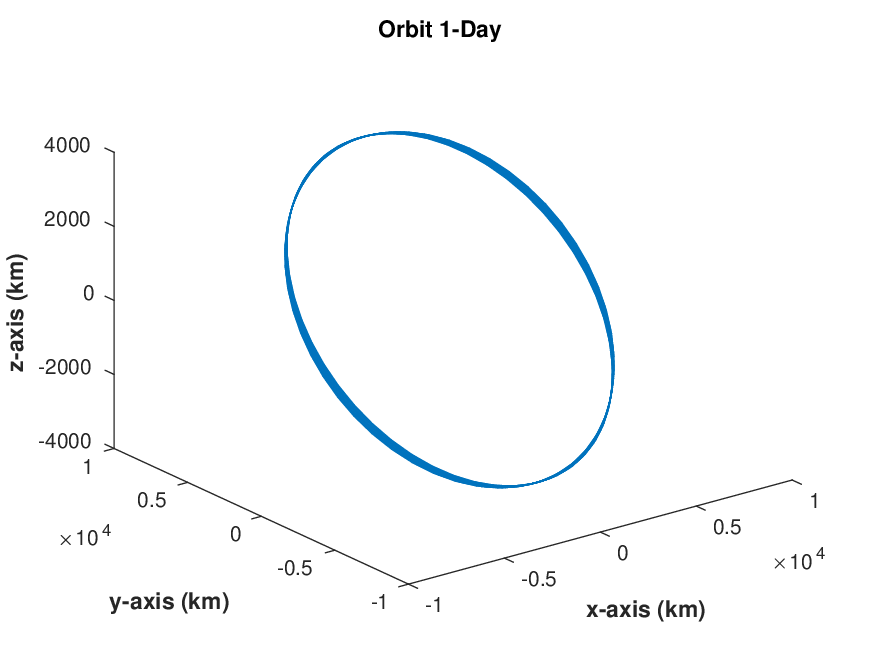} & \includegraphics[width=6cm,height=6cm]{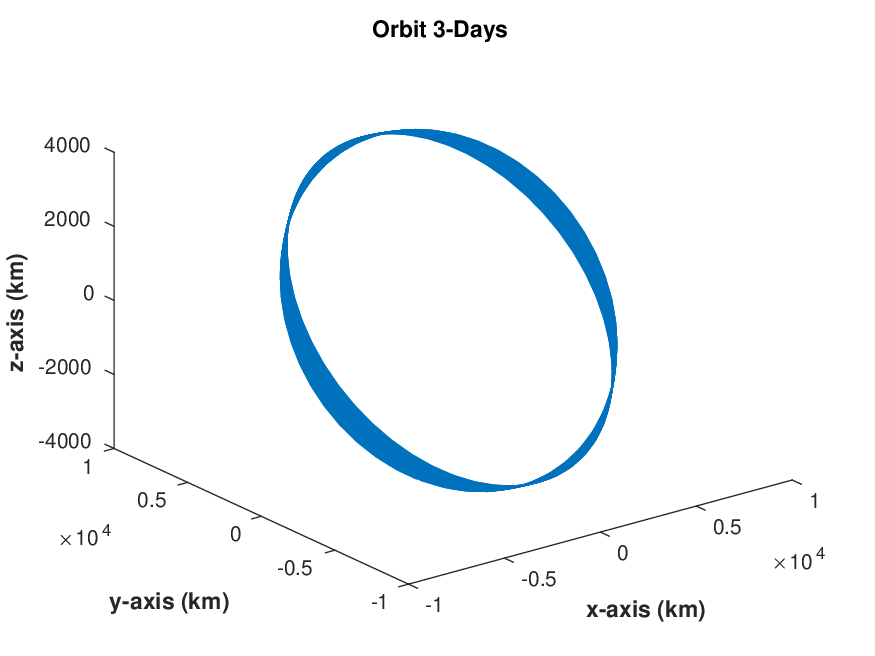} & \includegraphics[width=6cm,height=6cm]{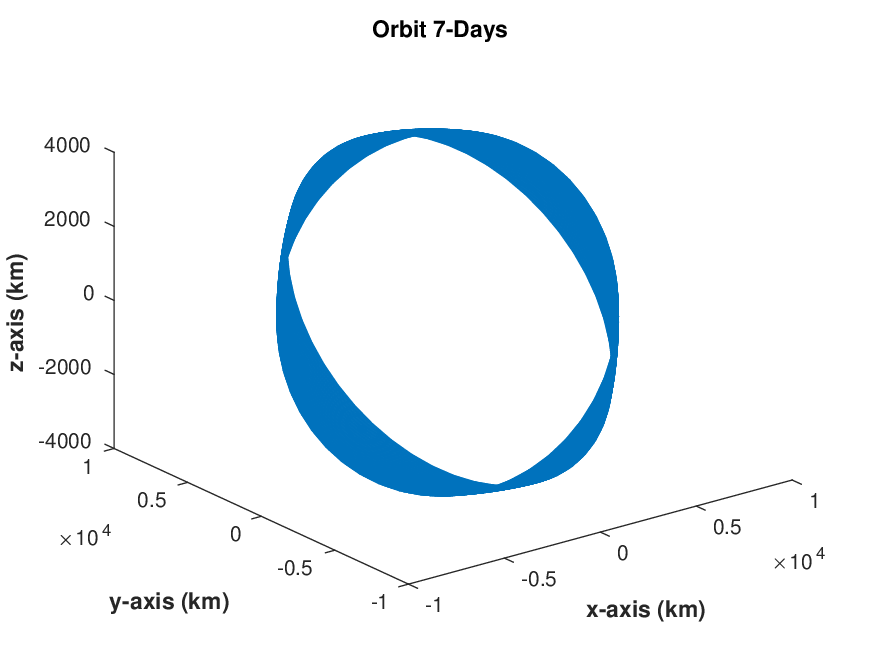}
\end{tabular}
\end{center}
\caption{Orbit of Satellite 1 day, 3 days and 7 days from left to right}
\end{figure}

\noindent The graphs of angle of inclination, argument of perigee, eccentricity, longitude of ascending node, semi major axis and
true anomaly with initial position $\vec{r}_{0}=[0, -5888.9727, -3400]\;km$, initial velocity $\vec{v}_{0}=[7.8, 0, 0]\;km/sec$ 
and ballistic coefficient $B^{*}=0.096\;m^{2}/kg;$ for 1 day, 3 days and 7 days are shown in figure 2, figure 3 and figure 4 respectively.

\pagebreak

\begin{figure}[H]
\begin{center}
\begin{tabular}{cc}
\includegraphics[width=6.5cm,height=6.5cm]{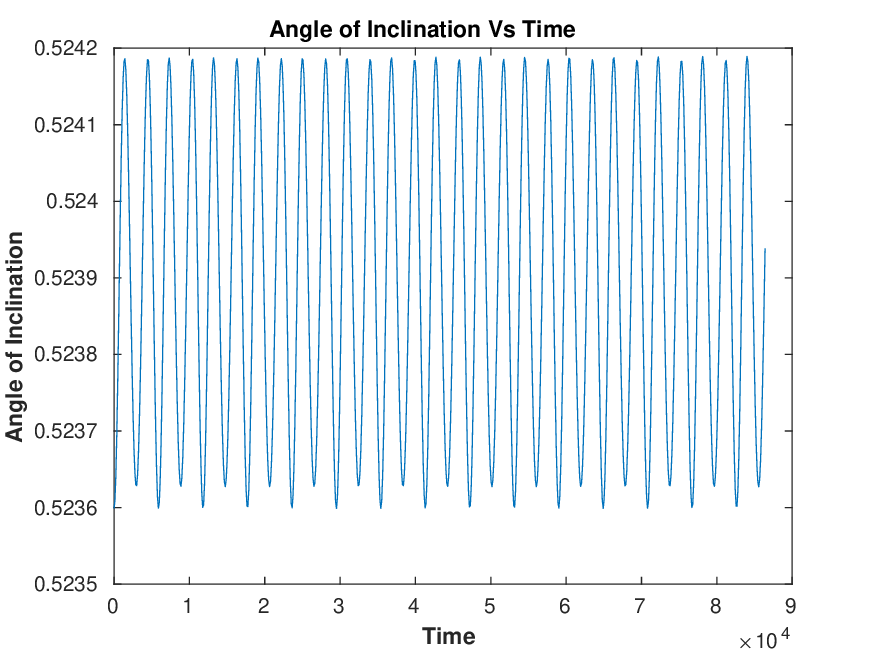} & \includegraphics[width=6.5cm,height=6.5cm]{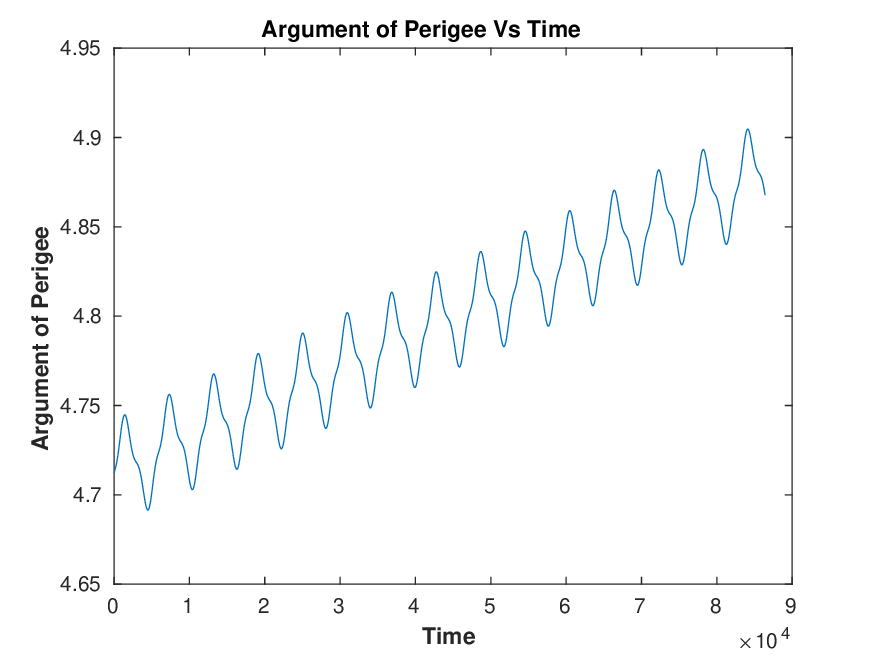} \\ 
\includegraphics[width=6.5cm,height=6.5cm]{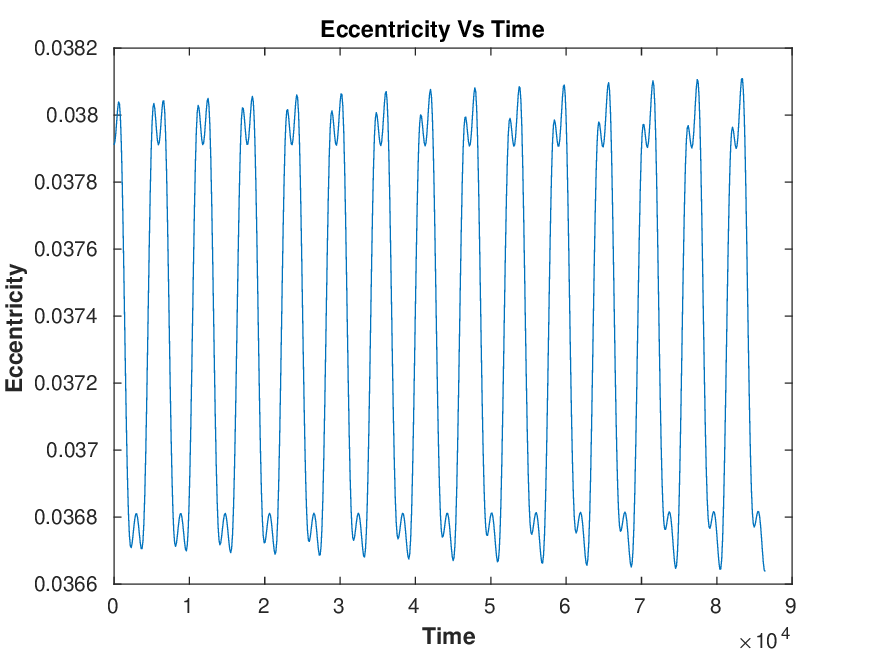} & \includegraphics[width=6.5cm,height=6.5cm]{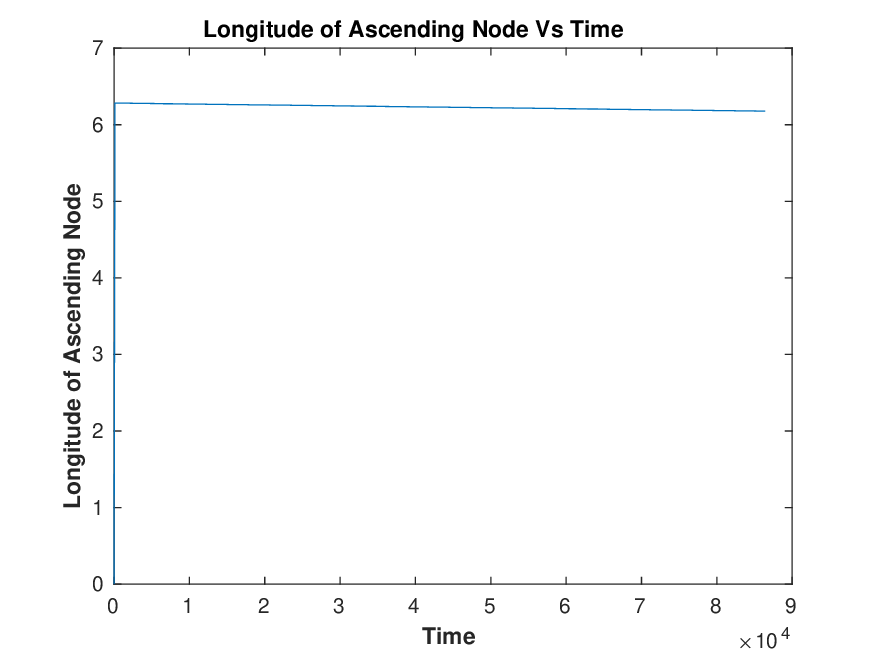} \\
\includegraphics[width=6.5cm,height=6.5cm]{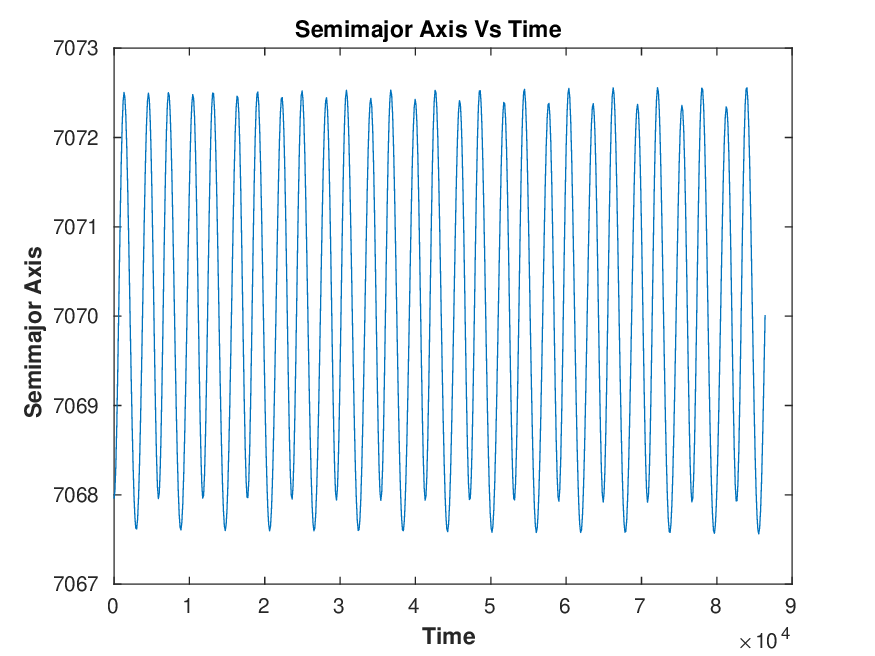} & \includegraphics[width=6.5cm,height=6.5cm]{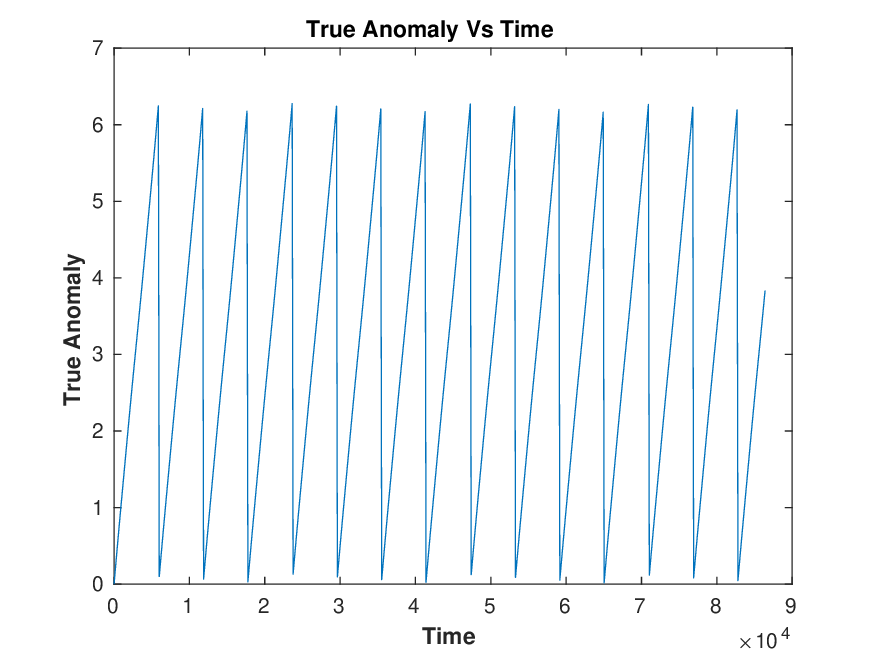}\\
\end{tabular}
\end{center}
\caption{Angle of inclination, argument of perigee, eccentricity, longitude of ascending node, semi major axis and
true anomaly with initial position $\vec{r}_{0}=[0, -5888.9727, -3400]\;km$, initial velocity $\vec{v}_{0}=[7.8, 0, 0]\;km/sec$ 
and $B^{*}=0.096\;m^{2}/Kg;$ for 1 day}
\end{figure}
\pagebreak

\begin{figure}[H]
\begin{center}
\begin{tabular}{cc}
\includegraphics[width=6.5cm,height=6.5cm]{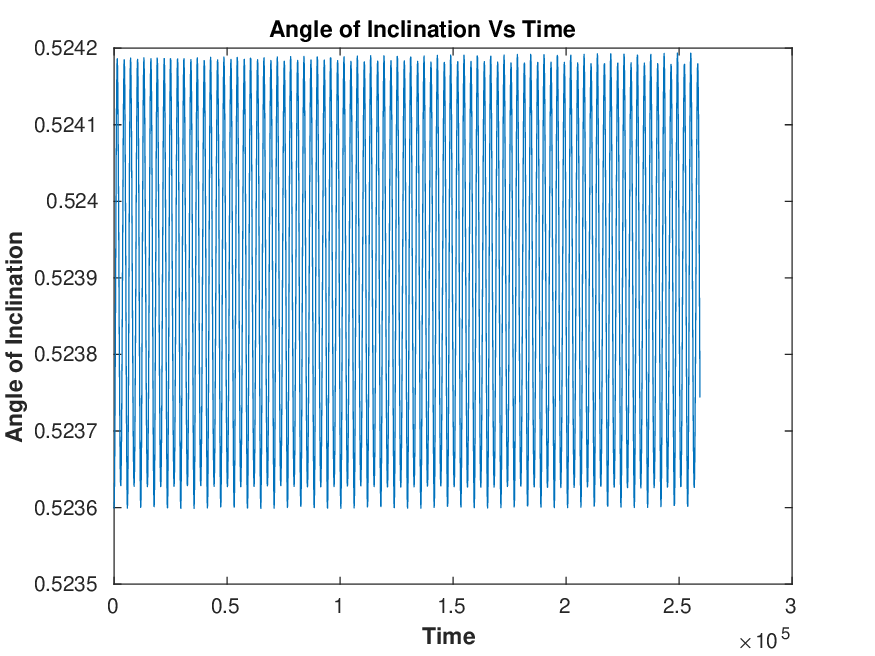} & \includegraphics[width=6.5cm,height=6.5cm]{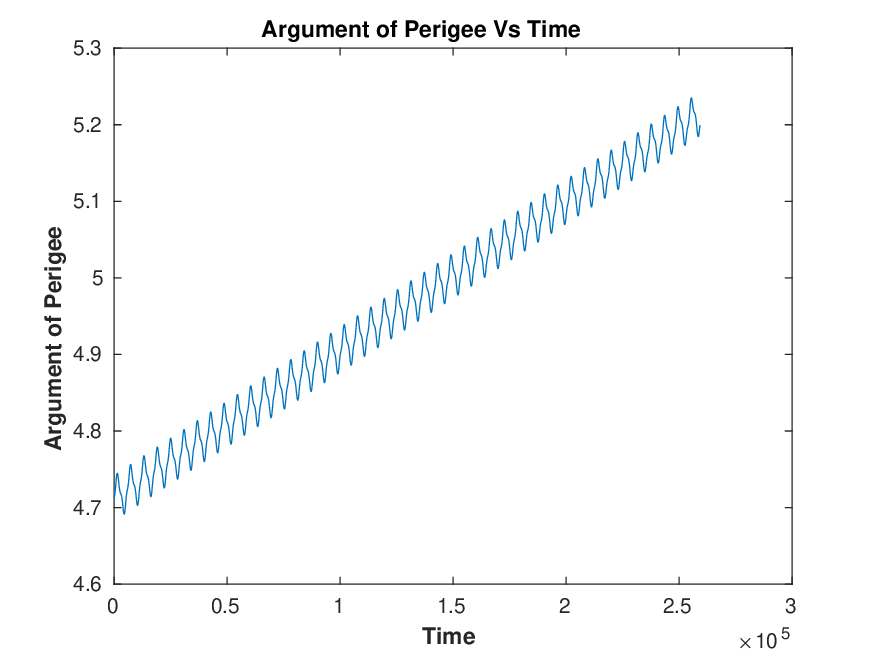} \\ 
\includegraphics[width=6.5cm,height=6.5cm]{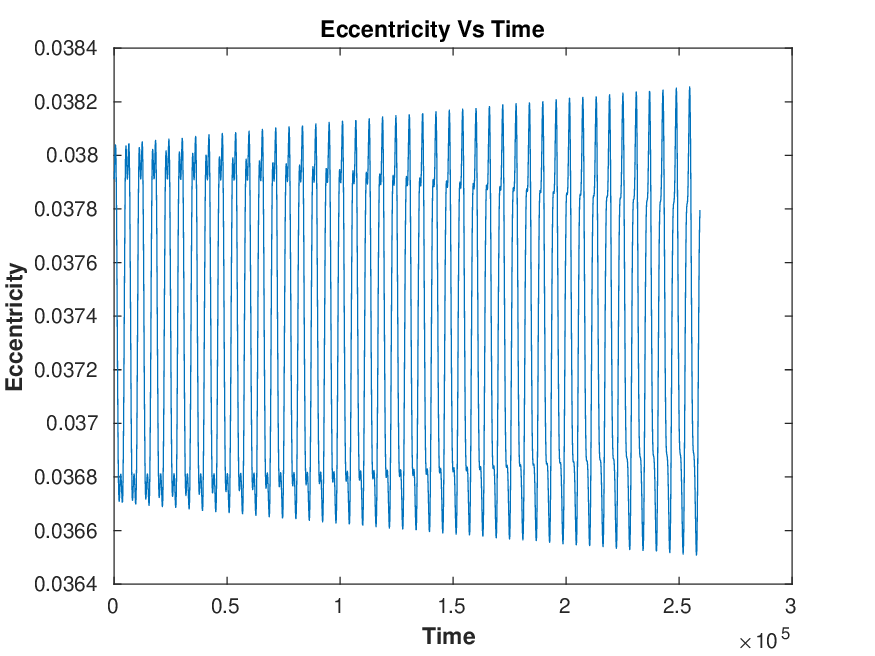} & \includegraphics[width=6.5cm,height=6.5cm]{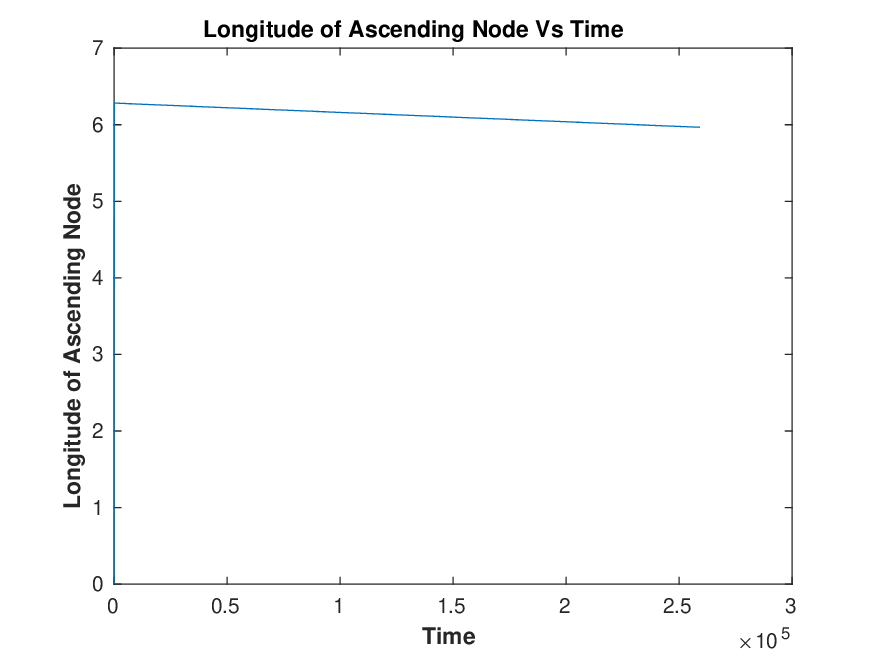} \\
\includegraphics[width=6.5cm,height=6.5cm]{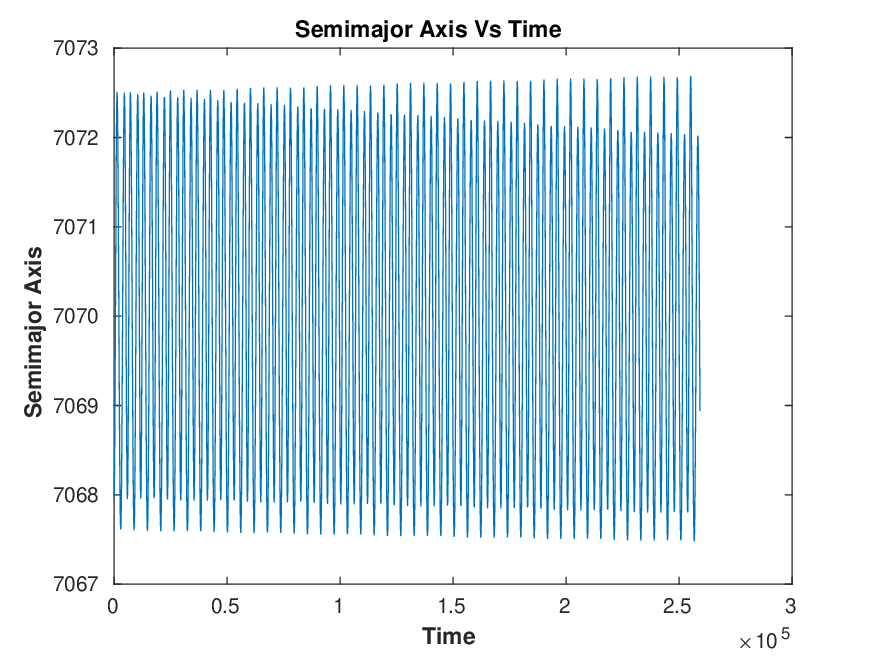} & \includegraphics[width=6.5cm,height=6.5cm]{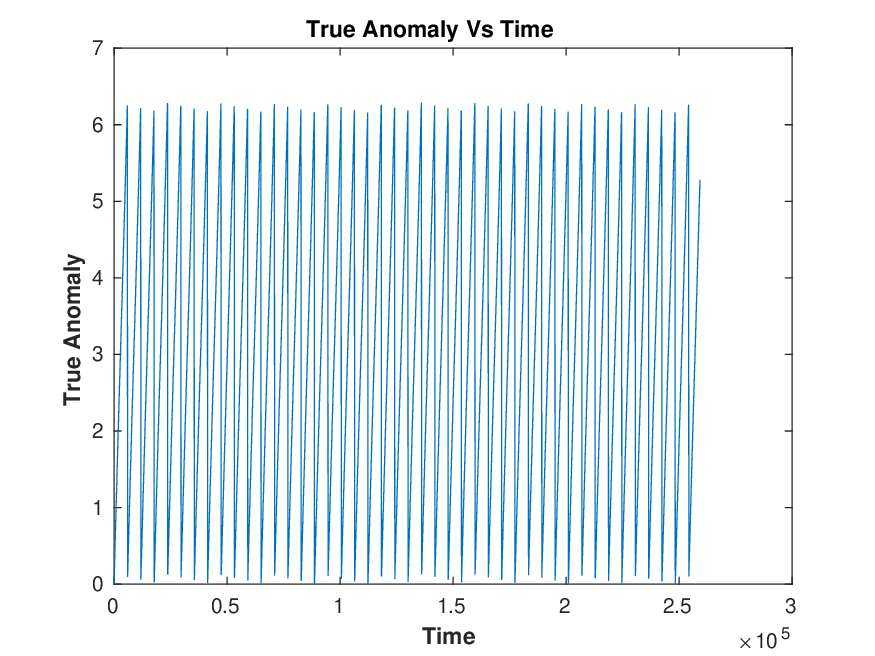}\\
\end{tabular}
\end{center}
\caption{Angle of inclination, argument of perigee, eccentricity, longitude of ascending node, semi major axis and
true anomaly with initial position $\vec{r}_{0}=[0, -5888.9727, -3400]\;km$, initial velocity $\vec{v}_{0}=[7.8, 0, 0]\;km/sec$ 
and ballistic coefficient $B^{*}=0.096\;m^{2}/Kg;$ for 3 days}
\end{figure}
\pagebreak

\begin{figure}[H]
\begin{center}
\begin{tabular}{cc}
\includegraphics[width=6.5cm,height=6.5cm]{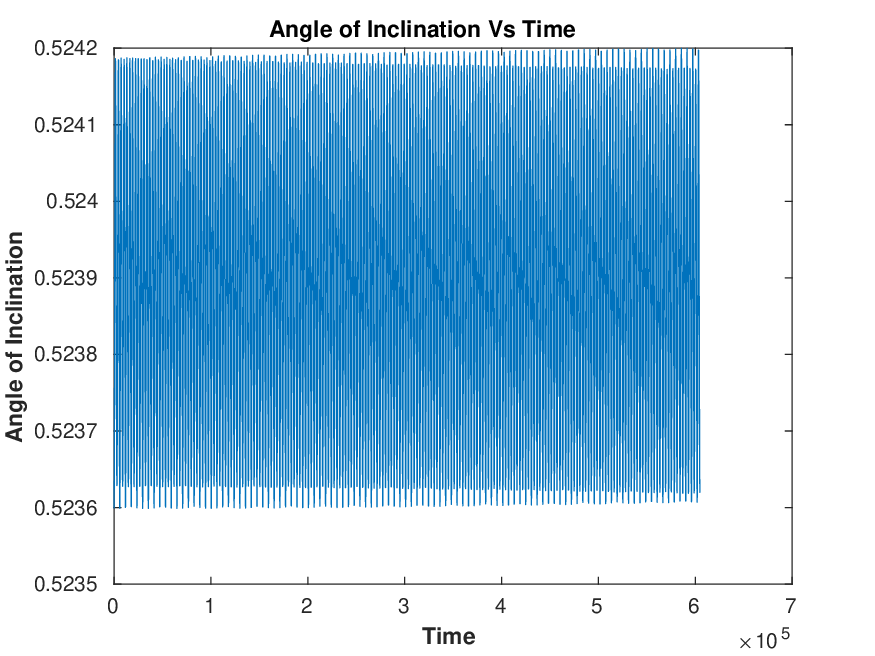} & \includegraphics[width=6.5cm,height=6.5cm]{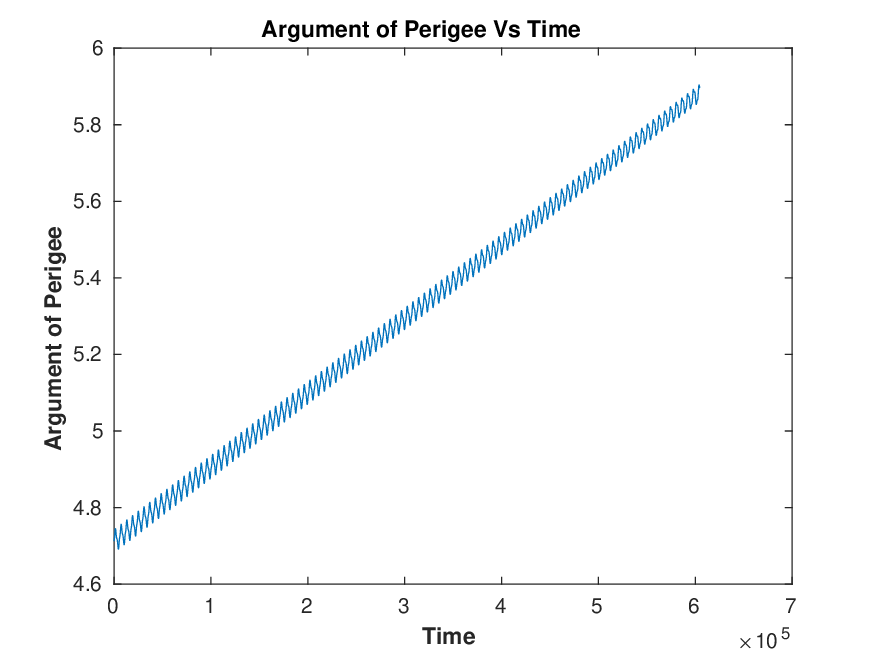} \\ 
\includegraphics[width=6.5cm,height=6.5cm]{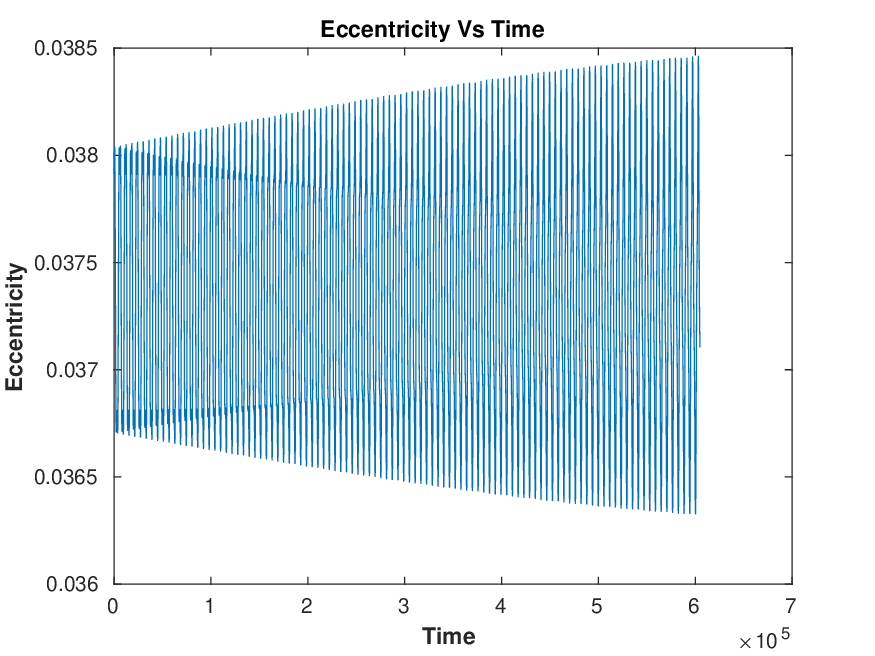} & \includegraphics[width=6.5cm,height=6.5cm]{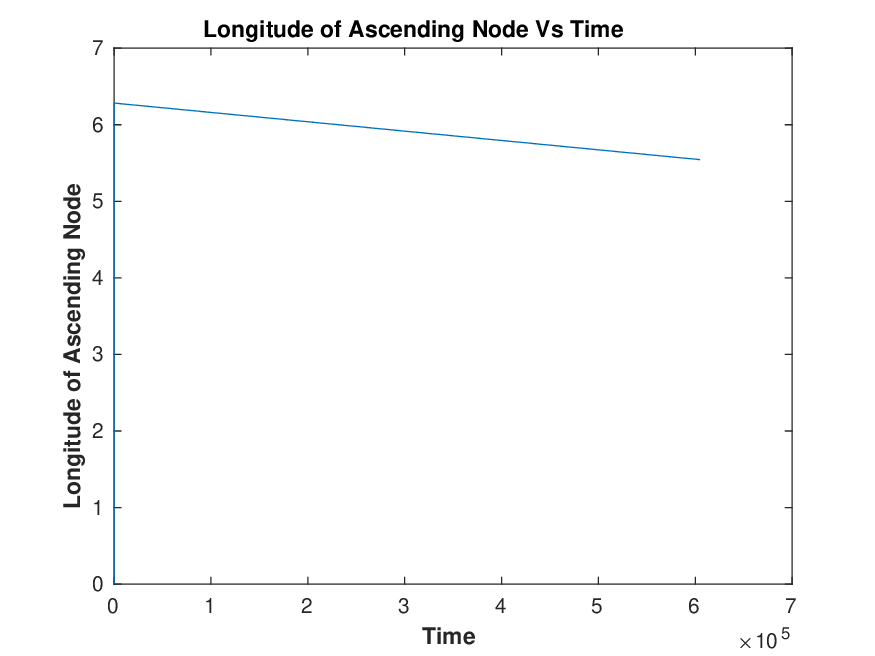} \\
\includegraphics[width=6.5cm,height=6.5cm]{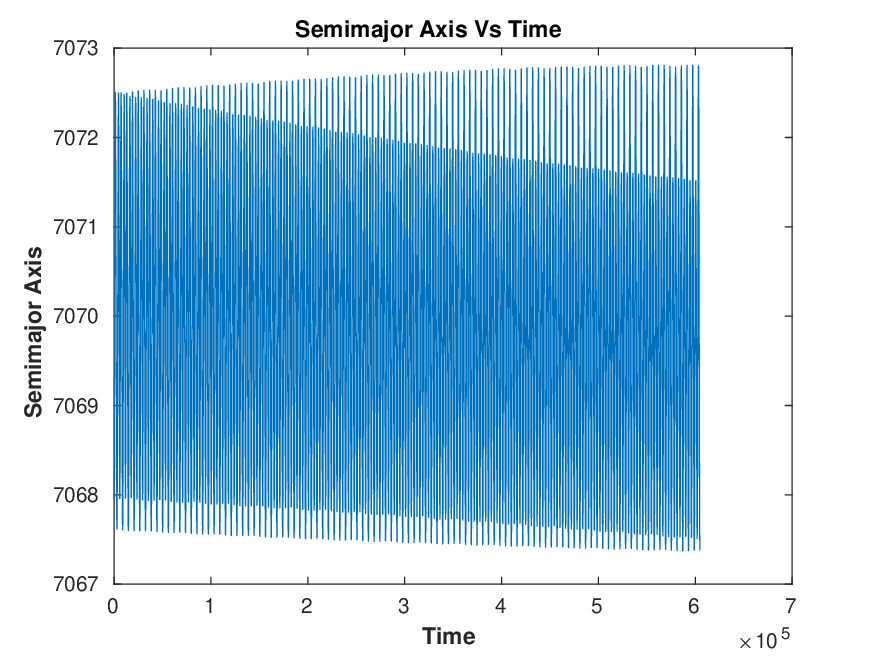} & \includegraphics[width=6.5cm,height=6.5cm]{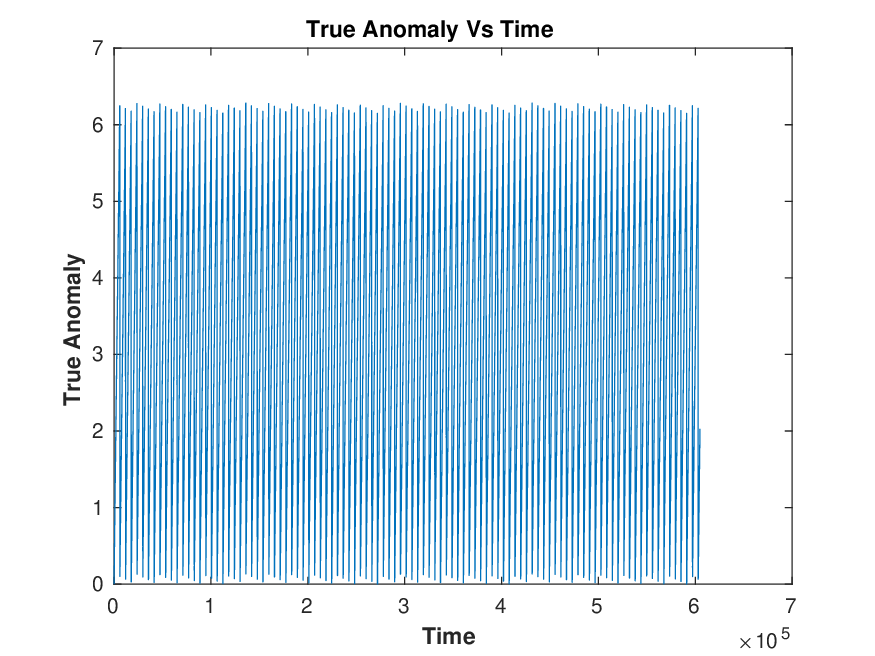}\\
\end{tabular}
\end{center}
\caption{Angle of inclination, argument of perigee, eccentricity, longitude of ascending node, semi major axis and
true anomaly with initial position $\vec{r}_{0}=[0, -5888.9727, -3400]\;km$, initial velocity $\vec{v}_{0}=[7.8, 0, 0]\;km/sec$ 
and ballistic coefficient $B^{*}=0.096\;m^{2}/Kg;$ for 7 days}
\end{figure}
\pagebreak

\section{Discussion and Concluding Remarks}
\noindent The equations governing motion of the satellite under the oblateness of Earth and atmospheric drag have been solved using R-K Gill over a short as well as long time duration.
In table-2 to table-4 minimum and maximum values of orbital elements over a different time period have been reported. From these
tables it can been seen that satellite will sustain for longer time in low Earth orbit if initial velocity is $\vec{v}_{0}=[7.8, 0, 0]\;km/sec$,
 initial position $\vec{r}_{0}=[0, -5888.9727, -3400]\;km$ and ballistic coefficient $B^{*}=0.096\;m^{2}/kg;$. From figure 1, it can be seen that even for a 
shorter period of time (1 day, 3 days and 7 days), oblateness of earth and atmospheric drag effects the orbit of the satellite. 
The salient features of solution of equation of motion under oblateness of Earth and Atmospheric drag are:
\begin{itemize}
	\item[1.] The choice of initial position and velocities are such that initially satellite's position is in $YZ$ plane and 		  velocities are applied in X-direction.
	\item[2.] The variation of argument of perigee is almost linear and increasing.
	\item[3.] The variation of logitude of ascending node is almost linear and decreasing.
	\item[4.] The eccentricity increases and then decreases over a longer time.
	\item[5.] The true anomaly varies between 0 to 6.2832.
	\item[4.] There is significant decline in semi-major axis over a long time duration.
	\item[5.] For initial position $\vec{r}_{0}=[0, -5888.9727, -3400]\;km$, initial velocity $\vec{v}_{0}=[7.8, 0, 0]\;km/sec$
and ballistic coefficient $B^{*}=0.096\;m^{2}/kg;$ satellite collapses on Earth after $9150.111111111111$ days, the height of 
satellite from surface of Earth on $9150^{th}$ day is approximately $43\;km$.
\end{itemize}
\noindent The particular care have been taken for step size of numerical integration in order to have stability for R-K Gill method. The analysis suggest that with mentioned initial position and initial velocities the maximum time the satellite can survive is $9150.111111111111$ days under the oblateness of Earth and Atmospheric drag.
\pagebreak

\end{document}